\documentclass[12pt]{article}
\pdfoutput=1
\usepackage{putex}

\usepackage{amsmath} % AMS Math Package
\usepackage{amsthm} % Theorem Formatting
\usepackage{amssymb}	% Math symbols such as \mathbb
\usepackage{graphicx} % Allows for eps images
\usepackage{multicol} % Allows for multiple columns

\usepackage[dvips,letterpaper,margin=1in,bottom=1.25in]{geometry}

\usepackage{float}
\usepackage{hyperref}
\usepackage{color}
\usepackage[obeyFinal]{todonotes}
\usepackage{float}
\usepackage{subcaption}
\usepackage{subfiles}
\usepackage{url}
\usepackage{indentfirst}
\usepackage{physics}

\usepackage[utf8]{inputenc}

\usepackage[nottoc,notlot,notlof]{tocbibind}

\usepackage{graphicx}

\def\F{{\cal F}}
\def\t{\tau}
\def\Vir{{\rm Vir}}
\def\Z{\mathbb{Z}}

\def\vac{{\rm vac}}
\def\half{{1\o 2}}
\def\N{{\cal N}}

\def\o{\over}

\def\0{{(0)}}
\def\1{{(1)}}
\def\2{{(2)}}
\def\3{{(3)}}
\def\4{{(4)}}

\def\Z{\mathbb{Z}}

\def\vac{\text{vac}}

\def\eqr{\eqref}
\def\sec{\section}

\def\l{\lambda}
\def\vs{\vskip .1 in}

\def\a{\alpha}
\def\rar{\rightarrow}

\def\la{\langle}
\def\ra{\rangle}
\def\O{{\cal O}}

\def\i{\infty}

\def\ssec{\subsection}
\def\sssec{\subsubsection}
\def\sec{\section}

\def\vs{\vskip .1 in}
\def\D{\Delta}

\newcommand{\e}[2] {\begin{equation} \label{#1} #2 \end{equation}}
\newcommand{\es}[2] {\begin{equation} \label{#1} \begin{split} #2 \end{split} \end{equation}}

\def\Vir{{\rm Vir}}

\def\0{{(0)}}
\def\1{{(1)}}
\def\2{{(2)}}
\def\3{{(3)}}
\def\4{{(4)}}

\def\vac{\text{vac}}

\def\eqr{\eqref}
\def\sec{\section}

\def\l{\lambda}
\def\vs{\vskip .1 in}

\def\a{\alpha}
\def\rar{\rightarrow}

\def\la{\langle}
\def\ra{\rangle}
\newcommand{\floor}[1]{\lfloor #1 \rfloor}
\newcommand{\W}{\mathcal{W}}
\newcommand{\bh}{\bar{h}}
\newcommand{\bz}{\bar{z}}

\newcommand{\btau}{\bar{\tau}}
\newcommand{\p}{\partial}
\definecolor{darkgreen}{rgb}{0,0.5,0}
\definecolor{darkpuce}{rgb}{0.5,0,0.5}

\newcommand{\be}{\begin{equation}}
\newcommand{\ee}{\end{equation}}

\graphicspath{{./Figures/}}

\numberwithin{equation}{section}

\begin{document}
	
\institution{cornell}{Department of Physics, Cornell University, Ithaca, NY, USA}
\institution{mcgill}{Department of Physics, McGill University, Montr\'eal, QC, Canada}
\institution{PU}{Department of Physics, Princeton University, Princeton, NJ, USA}

\title{Constraints on Higher Spin CFT$_2$}
\authors{Nima Afkhami-Jeddi\worksat{\cornell}, Kale Colville\worksat{\mcgill}, Thomas Hartman\worksat{\cornell}, \\
\vskip .1 in 
Alexander Maloney\worksat{\mcgill}, Eric Perlmutter\worksat{\PU}}
\abstract
{
We derive constraints on two-dimensional conformal field theories with higher spin symmetry due to unitarity, modular invariance, and causality. We focus on CFTs with $\W_N$ symmetry in the ``irrational" regime, where $c>N-1$ and the theories have an infinite number of higher-spin primaries. The most powerful constraints come from positivity of the Kac matrix, which (unlike the Virasoro case) is non-trivial even when $c>N-1$.  This places a lower bound on the dimension of any non-vacuum higher-spin primary state, which is linear in the central charge.  At large $c$, this implies that the dual holographic theories of gravity in AdS$_3$, if they exist, have no local, perturbative degrees of freedom in the semi-classical limit.}
\date{}
\maketitle
\setcounter{tocdepth}{2}
\tableofcontents
%
%\newpage

\section{Introduction and Discussion}

Conformal field theories are highly constrained by symmetry, but a complete classification of unitary CFTs remains out of reach. Prospects are better for two-dimensional CFTs, which possess an infinite algebra of local conformal symmetries \cite{Belavin:1984vu}.  Assuming no conserved currents other than the stress tensor, there are two qualitatively different classes of two-dimensional CFTs: rational CFTs with $c<1$ and irrational CFTs with $c>1$.  Rational CFTs, which possess a finite number of primary states, are integrable. They have been classified \cite{Cappelli:1986hf}, and the observables are completely computable using Virasoro Ward identities.  Irrational CFTs, which possess an infinite number of primary states, are more difficult to study.  The local conformal symmetries are not sufficient to render these theories completely integrable, so irrational CFTs are generically expected to exhibit many of the rich and complicated phenomena of their higher-dimensional cousins, including ergodicity and chaos (see for example \cite{Asplund:2015eha}). In this paper we will ask whether this general picture holds when, in addition to local conformal transformations, the CFTs possess higher spin symmetries, generated by conserved currents of spin $s>2$.

Theories with higher spin symmetry have been extensively investigated in two and higher dimensions.  In two dimensions the higher spin currents form an algebra of chiral operators -- a ${\cal W}$-algebra -- which is constrained by associativity of the operator product expansion (see \cite{Bilal:1991eu,Bouwknegt:1992wg} for early reviews).  Most past work has focused on rational theories, with central charge $c$ less than the number $N_{currents}$ of conserved currents. For Virasoro, $N_{currents} = 1$, while more generally for the $\W_N$ algebra, which is generated by currents of integer spins $s=2\ldots N$, one has $N_{currents} = N-1$. When $c<N_{currents}$ it is possible to construct higher spin CFTs with a finite number of primaries, in analogy with Virasoro minimal models with $c<1$. We will be interested in theories with $c>N_{currents}$. These are the higher spin analogue of the irrational CFTs, in the sense that they necessarily contain an infinite number of primary operators. We will restrict our attention to unitary, modular invariant theories with a discrete spectrum and a unique, normalizable ground state; this means we are not, for example, considering Toda theories. Here and below, by primary operator we mean an operator which is primary with respect to the full chiral algebra, not just Virasoro. The semiclassical, holographic limit is $c \gg N_{currents}$.

These irrational higher-spin CFTs resemble their higher-dimensional cousins, where the constraints of higher spin symmetry have proven quite powerful.  For example, Maldacena and Zhiboedov \cite{MaldacenaZhiboedev} have shown that a three-dimensional CFT with a conserved current of spin $s>2$ must contain an infinite tower of higher spin currents, whose correlation functions equal those of a free theory. Similar results apply in higher dimensions \cite{Boulanger:2013zza,Alba:2015upa}. At first sight this appears quite different from the situation in two dimensions, where it is possible to construct ${\cal W}$-algebras with only a finite number of generators.  
We will see, however, that this is somewhat misleading, and that -- with appropriate assumptions -- the unitarity constraints on irrational higher spin CFTs are similar to those of Maldacena and Zhiboedov, while perhaps also allowing for exotic higher spin theories with no higher-dimensional analogue.

We will focus primarily on the case of $\W_N$ algebras, although many of our results can be extended to other ${\cal W}$-algebras. For $N>2$, these algebras are non-linear. In Virasoro (i.e. $W_2$) theories without additional higher spin symmetries, the starting point for the construction of rational CFTs is the computation of the Kac matrix, the matrix of inner products of states in a lowest weight module 
of dimension $h$.\footnote{Here $h={(\Delta +J)/2}$ is the right-moving dimension of a state of scaling dimension $\Delta$ and spin $J$, and ${\bar h} = {(\Delta - J)/2}$ the corresponding left-moving dimension.   Since the symmetry algebra factorizes into left- and right-moving pieces, our results for the unitarity bounds will be stated directly in terms of $h$, with similar results for ${\bar h}$ implied.}  In irrational $(c>1)$ theories, the eigenvalues of the Kac matrix are positive definite whenever $h>0$.  Thus in the Virasoro case, the simplest constraint of unitarity -- that all states have positive norm -- is essentially trivial.  

The central surprise of our paper is that in higher spin CFTs, positivity of the Kac matrix strongly constraints the spectrum.  For example, we will argue that in the $\W_N$ case every (non-identity) primary operator must have dimension
\begin{equation}\label{result}
h \ge h_{crit}  = {c-(N-1)\over 24}\left(1-  \frac{6\floor{\frac{N}{2}}}{N(N^2-1)}\right)~.
\end{equation}
All states with $h<h_{crit}$ are in the identity block of $\W_N$, i.e. they are (higher spin) descendants of the vacuum.   The bound applies even to uncharged primaries, which necessarily have descendants carrying $\W$-charge. This is to be contrasted with the typical result of the conformal or modular bootstrap, which is an {\it upper} bound on the dimension of the lightest non-identity operator. Charged primaries obey stronger bounds. We will give a complete derivation of \eqref{result} for $N=3$, and numerical evidence for $N \leq 6$; for higher $N$, this bound is a conjecture based on the structure of null states.

The first thing to note about \eqref{result} is that it is trivial in the rational case, $c<N-1$.  Indeed, the bound becomes stronger as $c$ is increased. Our $\W_N$ result \eqref{result} follows directly from the unitarity of the Kac matrix at level one, and it may be possible to obtain higher bounds at higher levels (though we suspect not at large $c$; see appendix \ref{app:w3level2}). 
  Irrational $\W_N$ CFTs, if they exist, are peculiar indeed: for instance, above the value of $c$ defined by the equation $h_{crit}(c_{*})=1$, a CFT with $\W_N\times \W_N$ symmetry is isolated, admitting no Lorentz-invariant marginal deformations. For $\W_3$, the critical central charge is $c_*=34$. We also provide a simple argument from modular invariance that every CFT has states with $h$ below a finite `twist gap'. Together with \eqr{result}, we now have upper and lower bounds on $h_1$, the weight of the lowest-twist non-trivial primary in a $\W_N$ CFT:
 \be\label{twistbound}
h_{crit}
 \le
h_1
\le
 {c-({N-1})\over 24}~.
 \ee
 This becomes especially tight as $N\to \infty$, where the upper and lower bounds coincide.  Thus in the large $N$ limit, with $c>N-1$, the lowest-twist state must have weight  (infinitesimally close to) $h_1 = (c-N+1)/24$. (This state may have large $\bar{h}_1$, so in particular it may have scaling dimension  $\D > (c-N+1)/12$, the threshold for the lightest black hole.)
 
Specializing to the case of $\W_3$, we obtain further constraints on the spectrum by applying modular bootstrap techniques \cite{Hellerman:2009bu, Friedan:2013cba, Qualls:2013eha, Benjamin:2016fhe, Collier:2016cls, Apolo:2017xip}, not to the partition function, but to the torus two-point function of $W_0$, the zero mode of the spin-3 current. This relies on the results of \cite{Iles:2013jha, Iles:2014gra}, which computed the modular transformation properties of this and similar objects. The results are bounds on the dimension and charge of primary operators that are independent of -- and indeed, have non-trivial overlap with -- the Kac matrix result \eqr{result}. We set up the algorithm for numerics, and study the linear functional analytically at large $c$. The main results can be found in and around Figures \ref{fig1}--\ref{fig2}.  Similar results can be obtained at small $c$ using numerical bootstrap.

Let us explore the consequences of \eqr{result} at large $c$. In \cite{Perlmutter:2016pkf} it was shown that the chaos bound on four-point functions forbids the existence of any non-trivial primary with dimension that remains finite in the limit $c\to\infty$ with $N$ fixed. We will provide a new argument, similar to that of \cite{Perlmutter:2016pkf}, that rules out light operators at large $c$ based on the lightcone limit of four-point functions, rather than the Regge limit, and makes fewer assumptions about the spectrum. For the $\W_N$ algebra, this dynamical argument is not necessary because the result is weaker than the unitarity of the Kac matrix \eqref{result}. The advantage is that unlike \eqref{result}, which is part conjecture, the causality argument applies to any higher spin chiral algebra.

Our results may be summarized by saying that every $\W_N$ CFT with a $c\to\infty$ limit is an extremal CFT, where we define ``extremal,''  slightly relaxing Witten's strict definition in the Virasoro context \cite{Witten:2007kt}, to include any family of CFTs with no primary operators whose dimension remains finite in the large $c$ limit.  Turning this around, the only way to construct a family of $\W_N$ CFTs with a finite dimension primary at large $c$ would be to scale $N$ linearly with $c$.  This result is in the spirit of Maldacena-Zhiboedov: in the large-$c$ limit, if there is a higher spin current and a finite dimension primary, then there must be an infinite tower of higher spin currents.

These large $c$ results have powerful implications for holography, as in \cite{Perlmutter:2016pkf}. A CFT${}_2$ with higher spin symmetry should be dual to a theory of quantum gravity in AdS$_3$ with higher spin gauge symmetry. In three dimensions, theories of higher spin gravity with a finite number of higher spin gauge fields may be constructed (classically) from non-compact Chern-Simons theories, following the example of ordinary gravity \cite{Witten:1988hc, Achucarro:1987vz, Campoleoni:2010zq}. These theories have no local degrees of freedom. In order to construct full-fledged holographic dual pairs, one must add matter to the bulk theory (e.g. to satisfy modular invariance of the CFT). A class of examples were proposed by Gaberdiel and Gopakumar \cite{Gaberdiel:2010pz}; these theories are rational, with  $c<N_{currents}$. Accordingly, the $c\to\infty$ limit requires simultaneously taking $N\to\infty$. This yields a bulk dual which is quite far from perturbative Einstein gravity. With an infinite number of gauge fields, this theory looks more like a theory of tensionless strings. Moreover, the duality proposal of \cite{Gaberdiel:2010pz} contains a plethora of light states, so it is unclear whether these theories have a holographic limit in the usual sense \cite{Chang:2011mz, Papadodimas:2011pf, Gaberdiel:2013cca, Castro:2010ce}. 

To instead obtain a higher spin dual which resembles Einstein gravity coupled to matter, one should take a semi-classical limit $c\to\infty$ with $N_{currents}$ fixed. Indeed, the loop-counting parameter in the bulk is $1/(c-N_{currents})$, not $1/c$. But in the $\W_N$ case, our bound (\ref{result}) constrains the resulting spectrum. In particular, all (non-identity) primaries have dimensions linear in $c$. In gravitational language, such primaries would be dual to particles with mass of order ${\cal O}(\ell_{AdS} /G)$.   A perturbative particle, on the other hand, such as that created by a quantum field of fixed mass, would have dimension that remains finite as $c\to\infty$.  Therefore, in the semi-classical limit the corresponding theory of gravity would have no local perturbative degrees of freedom.

Although much of this discussion has been phrased in terms of the $\W_N$ algebra, many of our results, with the exception of \eqref{result} and the modular bootstrap, naturally extend to any $\W$-algebra. In general, a theory is irrational, in the sense that it must have an infinite number of primaries, when $c > c_{currents}$ where $c_{currents}$ is the effective central charge of the vacuum representation.\footnote{That is, $\log \chi_\vac(\tau\rar 0) \sim {i\pi \o 12\tau} c_{currents}$, where $\chi_\vac$ is the vacuum character of $\W$.} Both the causality argument and the upper bound on the twist gap apply to any irrational $\W$-algebra, so the smallest $h$ must grow with $c$ and lie below $(c - c_{currents})/24$.

In all, the above considerations suggest that irrational higher spin CFTs should be difficult to construct, especially in the large $c$ limit.  Indeed, our initial goal was to prove that $\W_N$ theories with $c>N-1$ do not exist.  We have not managed to do so. However, the fact remains that there is not a single known example of a $\W_N$ CFT with $c>N-1$ and a normalizable ground state. Moreover, a scan of various examples in the literature reveals that all unitary CFTs in the irrational regime that have higher spin symmetries, with any chiral algebra, contain non-trivial null states in the vacuum module; we collect some of these examples in Appendix B. Based on this and the surprising strength of the constraints derived herein, a natural question about the space of higher spin CFT$_2$ is: 
\vs
\begin{quote}
\it  Does the vacuum module of a unitary, higher spin CFT$_2$ with finite central charge always have non-trivial null states?
\end{quote}

\noindent Perhaps so. A slightly narrower possibility is that freely generated algebras, such as $\W_N$, are only realized in unitary CFT if the theory is rational, with $c=c_{currents}$. This would be a stronger two-dimensional version of the Maldacena-Zhiboedov result, in which rational theories are playing the role of the free theories in three dimensions.
\vs

The rest of the paper is organized as follows. In Section 2, we derive the twist bound, which is the upper bound in \eqr{twistbound}. In Section 3, we derive the absence of light operators directly at large $c$ using the lightcone limit of four-point functions. These results apply to any finitely-generated $\W$-algebra. In Section 4, we relax to finite $c$ and study the Kac matrix of $\W_N$, deriving the bound \eqr{result} and providing more detailed plots of the allowed parameter space. In Section 5, we apply modular bootstrap techniques to traces over $\W_3$ zero modes, yielding constraints on the charge and dimension of non-identity operators in a $\W_3$ CFT. In Appendix A, we prove that the $\W_3$ unitarity constraints are saturated by the level-one Kac matrix for $c\leq 98$. In Appendix B, we present a partial accounting of known unitary CFTs with higher spin symmetries and their null states.

\section{Upper bound on the twist gap}
We begin with a simple argument that, in an irrational theory with $\W_N$ symmetry, there must be an accumulation point in the spectrum at
\begin{equation}\label{accum}
h = \frac{c-N_{currents}}{24} \ ,
\end{equation}
with $N_{currents} = N-1$. That is, there is an infinite number of states arbitrarily close to \eqref{accum}. This claim does not assume large $c$, and it is true and non-trivial even with just Virasoro symmetry.\footnote{This was previously reported and applied to the bootstrap in \cite{Collier:2016cls}.} It comes from a chiral limit of the Cardy formula \cite{Cardy:1986ie}. The states accumulating at \eqref{accum} have $\bar{h} \to \infty$; identical statements apply to the spectrum of $\bar{h}$ as $h \to \infty$. This is therefore an accumulation point in the twist spectrum,
\be\label{twistdef}
\mbox{twist} \equiv \Delta - J = 2\,  \mbox{min}(h,\bh) \ ,
\ee
 coming from high-energy states with large spin. In conjunction with lower bounds to be derived later in the paper, this implies a narrow window of twists in which operators must exist in any $\W_N$ CFT, as quoted in \eqref{twistbound}.

The strategy is similar to the lightcone bootstrap, initiated in \cite{Komargodski:2012ek,Fitzpatrick:2012yx}. To briefly summarize the main result of these papers, consider the four-point function $G(z,\bz)$ for a scalar operator of dimension $\Delta$, in a CFT in $d>2$, in Lorentzian signature where the cross-ratios $z$ and $\bz$ are independent real numbers. The crossing equation, $G(z,\bz) = G(1-z,1-\bz)$, is expanded in the double lightcone limit, $z \ll 1 - \bz \ll 1 $. In this limit, one channel is dominated by the disconnected product of two-point functions -- i.e., the identity conformal block -- and to reproduce this singular behavior in the other channel, there must be an infinite sum over operators with twist accumulating at $2\Delta$. 

To derive the accumulation point \eqref{accum}, we will apply similar logic to the 2d partition function, $Z(\tau,\btau)$. The crossing equation in this context is modular invariance:
\be\label{modinv}
Z(\tau,\btau) = Z(-1/\tau,-1/\btau) \ .
\ee
Modular invariance holds for independent complex $\tau$ and $\btau$ (see e.g. footnote 3 of \cite{Hartman:2014oaa}), so we may choose $\tau = \frac{i\beta_L}{2\pi}$, $\btau = \frac{i\beta_R}{2\pi}$, with $\beta_L$ and $\beta_R$ independent real numbers. This corresponds to the partition function with real temperature and angular potential, $Z = \mbox{Tr} e^{-\beta (H - \Omega J)} = \mbox{Tr}  e^{-\beta_L (h-c/24) - \beta_R (\bh-c/24)}$, and is analogous to the Lorentzian four-point function used above. The accumulation point \eqref{accum} is derived by solving \eqref{modinv} in the highly rotating, low temperature limit, $\beta_R \gg \frac{1}{\beta_L} \gg 1$. This is the partition function analogue of a double lightcone limit. (In fact they are identical if the partition function is viewed as a four-point function of twist operators \cite{Asplund:2015eha}.)

This result has a clear bulk interpretation, in both ordinary gravity and higher spin gravity. Operators with $h\approx {c-N_{currents}\o24}$ and parametrically larger $\bar h$ are dual to microstates of large extremal BTZ black holes. The fact that such operators must exist is a CFT derivation of the existence of such black holes, at arbitrary $c$. Said another way, the ``Cardy regime'' requires only one of the temperatures to be large.

\bigskip
\noindent\textit{Derivation:}
\vs

In the limit $\beta_R \to \infty$, with $\beta_L$ held fixed, the l.h.s. of \eqref{modinv} is dominated by the vacuum character,
\begin{align}
Z(\tau,\btau) &= \sum_{h,\bh} \rho(h,\bh) \chi_h(\tau)\chi_{\bh}(\btau)\\
&\approx   \chi_{\vac}(\tau) e^{\beta_R c/24} \ ,\label{zzero}
\end{align}
where $\chi_h(\tau)$ is the character of the $\W_N$ algebra with weight $h$, and $\chi_{\vac}(\t)$ is its vacuum character. Assuming $c > N -1$, there are no non-trivial null states in the vacuum representation, so the vacuum character simply counts integer partitions of the current modes acting on the vacuum state:
\be  
\chi_{\vac}(\tau)  = q^{-c/24}\prod_{s=2}^N \prod_{n=s}^\infty (1-q^n)^{-1} \qquad (q \equiv e^{2\pi i \tau}) \ .
\ee
Next, send $\beta_L \to 0$ ($q \to 1$).  In this limit, the $\W_N$ vacuum character can be evaluated by a saddlepoint approximation (or by the Cardy formula applied to a theory of $N-1$ free chiral bosons) and behaves as
\be
\chi_\vac(\tau) \approx (q')^{-(N-1)/24}  \qquad (q' \equiv e^{-2\pi i/\tau}) \ .
\ee
Therefore, modular invariance in the limit $\beta_R \gg \frac{1}{\beta_L} \gg 1$ requires
\be\label{zff}
\exp\left( \frac{4\pi^2}{\beta_L} \frac{N-1}{24} + \beta_R \frac{c}{24} \right) \approx \sum_{h,\bh} \rho(h,\bh) \exp\left(-\frac{4\pi^2}{\beta_L} (h-c/24) - \frac{4\pi^2}{\beta_R}(\bh-c/24)\right)
\ee
(On the r.h.s we have used the density of primaries $\rho(h,\bh)$ to approximate the density of all states, which is valid in this limit because $N$ is held fixed and finite.)

The $\beta_L$-dependence on the left of \eqref{zff} must be reproduced on the right by states with $h$ given by \eqref{accum}, or more accurately, by states with $h$ arbitrarily close to \eqref{accum}. Furthermore, the left is singular as $\beta_R \to \infty$, while each character on the right is regular in this limit, so this term can only be reproduced on the right by the asymptotics of the infinite sum as $\bh \to \infty$.  The conclusion is that we must have an accumulation point in the spectrum of $h$ at \eqref{accum} as $\bh \to \infty$. In particular, inserting the factor of 2 from \eqref{twistdef}, the twist gap can be no larger than twice \eqref{accum}.
\vs
This result and derivation generalize straightforwardly to other algebras with finite $N_{currents}$. One simply replaces $N-1$ with $c_{currents}$, defined by the asymptotic growth of the vacuum character as $\chi_\vac(\tau) \approx (q')^{-c_{currents}/24}$. In the absence of null states in the algebra, $c_{currents} = N_{currents}$.

Note that the assumption $c > c_{currents}$ was crucial in this derivation. Otherwise, the vacuum in one channel can be reproduced by an individual state in the other channel, and there is no need for an infinite sum. This is how modular invariance is satisfied by a finite spectrum of primaries in a rational CFT.

\section{Large $c$ constraints}\label{s:causality}
Our next goal is to prove that a large-$c$, irrational 2d CFT, with higher spin symmetry that couples to at least one light primary, must have $N_{currents} = \infty$. This is a weaker, 2d version of the Maldacena-Zhiboedov theorem, as discussed in the Introduction. In this section, `light' means that the conformal dimension $h$, and higher spin charges, are fixed in the large-$c$ limit, $h \sim O(c^0)$ (but $h\gg 1$ is allowed). The result derived here is weaker than the more general bound derived from the Kac matrix below, which allows for a range of $h$ scaling with $c$, but it has two advantages. First, it is more similar in spirit to the derivations in higher dimensions, and gives some physical intuition for why higher spin symmetry is disallowed in the semiclassical limit. Second, it does not rely on the assumption about null states that we will invoke in section \ref{iii} below to extend the Kac results to $N > 6$, and applies to any higher spin algebra.

\subsection{Causality}

The first argument is a slight modification of \cite{Perlmutter:2016pkf}, where the same result was derived under the extra assumption of a sparse spectrum of low-dimension operators. In \cite{Perlmutter:2016pkf}, the strategy was to compute the contribution of higher spin exchange to the four-point function of a light operator in the Regge limit, then to apply the chaos bound of Maldacena, Shenker, and Stanford \cite{Maldacena:2015waa}. Here, following \cite{Hartman:2015lfa}, we will repeat the argument in the lightcone limit, rather than the Regge limit. The lightcone limit guarantees that conserved currents dominate the conformal block expansion, regardless of what other light states appear in the spectrum, so avoids any assumption about sparseness. The method is otherwise identical to \cite{Perlmutter:2016pkf}, so we will be brief. 

The $\W_N$ algebra is generated by a tower of spin-$s$ currents with $s=2,3,\ldots N$, each of which admits a Laurent expansion in modes $W^s_m$, with $m\in\Z$. Primary states of $\W_N$ are labeled by higher spin charges,
\begin{equation}
\ket{h,q_3,..., q_N },
\end{equation}
where
\e{}{L_0 \ket{h,q_3,..., q_N} = h \ket{h,q_3,..., q_N} \ , ~~~
W^{s}_0 \ket{h,q_3,..., q_N} = q_s \ket{h,q_3,..., q_N}  \ .}
(We are not considering chiral CFTs, so there are also anti-holomorphic labels which are suppressed and play no role in this paper.) We take a generic $\W_N$-primary operator $\O$ to have these charges, via the state-operator correspondence. The four-point function of ${\cal O}$ can be expanded in $\W_N$ conformal blocks,
\be
G(z,\bz) = \sum_{h,\bh} c_{h,\bh} \F_h(z) \F_{\bh }(\bz) \ .
\ee
where we suppress the dependence of the blocks on the $q_i$. In the lightcone limit, $\bz \to 0$ with $z$ held fixed, the sum is dominated by operators with zero twist. These are precisely the conserved currents of the theory, which are all descendants of the vacuum, so only the vacuum block survives:
\be\label{gofz}
G(z,\bz) \approx \bz^{-2\bh_O} \F_{\vac}(z) \ .
\ee
In the large-$c$ limit, the vacuum block is a sum of hypergeometric functions, one for each conserved current:\footnote{In general, the vacuum conformal block is the projection
\be\label{genvac}
\F_{\vac}(z) \equiv \sum_{p,q} \langle {\cal O}(0) {\cal O}(z) Q_p|0\rangle \langle 0|Q^\dagger_q {\cal O}(1) {\cal O}(\infty)\rangle (K_{pq})^{-1}
\ee
where the sum is over all states created by acting on the vacuum with any combination of current modes: 
$Q_p \equiv W^{s_1}_{-n_1} W^{s_2}_{-n_2} \dots$ with $n_i \geq s_i$.
$W^s_m$ denote modes of the spin-$s$ current $W^s(z)$, and $K_{p q} \equiv \langle Q^{\dagger}_p Q_q\rangle$ is the vacuum Kac matrix. In the limit $c \to \infty$ with the external weights held fixed, the leading contributions to the vacuum block come from states with only a single mode:
\be\label{remst}
W^{s}_{-m}|0\rangle , \qquad s=2\dots N , \ m\geq s \ .
\ee
These states have norms $\sim c$, while all other states have norms scaling as higher powers of $c$, so they are subleading due to the $K_{pq}^{-1}$ term in \eqref{genvac}.
In other words, only the single-trace operators, $\p^m T(z) , \p^n W^3(z), \dots$ survive the large-$c$ limit. These states are simply the conserved currents and their $SL(2)$ descendants.  It follows that the large-$c$ vacuum block is a sum of $SL(2)$ blocks, which are given by the hypergeometric functions written in \eqref{wnvac} \cite{Ferrara:1974ny, Fateev:2011qa}.}
\be\label{wnvac}
\F_{\vac}(z) = z^{-2h_O}\sum_{s =2}^N \frac{q_s^2}{c {\cal N}_s} z^s {}_2F_1(s,s,2s;z) \ ,
\ee
where we have normalized the currents such that $\la W^s_{s}W^s_{-s}\ra = c {\cal N}_s$.

To rule out higher spin theories in this context, we will apply the chaos bound of \cite{Maldacena:2015waa}, adapted to lightcone kinematics in \cite{Hartman:2015lfa}. The relevant results of these papers are summarized as follows. Define the normalized four-point function
\be
G_\eta(z)  = z^{2h_O}(\eta  z)^{2\bh_O} G(z,\eta z) \ .
\ee
Fix $\eta \ll 1$, which is the lightcone limit. This function has a singularity at $z = 1$. Define the correlator on the `2nd sheet', $\hat{G}_\eta(z)$, by analytically continuing around this singularity, along the path
\be
(1 - z) \to (1 - z)e^{-2\pi i} \ .
\ee
Then the bound states that $|\hat{G}_\eta(z)-1|$ can grow no faster than $1/|z|$ as $|z| \to 0$. This follows from analyticity, which in turn follows from causality in the sense that $\langle {\cal O} [ {\cal O}, {\cal O}]{\cal O}\rangle$ must vanish in the appropriate regime.

Returning to our argument, the correlator \eqref{gofz}, using \eqref{wnvac}, behaves in this regime as 
\be
\hat{G}_\eta(z) -1 \sim z^{-(N-1)} \ .
\ee
Therefore it violates the chaos bound for any finite $N>2$. The conclusion is that if there is any higher spin current coupling to a light primary, then there must be an infinite number of such currents. If the sum in \eqref{wnvac} is infinite then the behavior near the origin can be softened. Although we have phrased this argument in the language of the $\W_N$ algebra for definiteness, this was not necessary, and it applies to any theory with higher spin conserved currents.

The Maldacena-Zhiboedov theorem in higher dimensions has a second part, which states that the correlators of the infinite tower of higher spin currents are necessarily those of a free theory. We have not derived any analogue of this statement in 2d. We have only shown that the higher spin currents must exist. It would be interesting to show that they have fixed correlators as well. A natural conjecture would be that the currents must reproduce the ${\cal W}_\infty[\lambda]$ algebra \cite{Gaberdiel:2011wb}.

\subsection{Unitarity}
The above argument used a dynamical constraint on four-point functions, and applied only to higher-spin charged operators $\O$.  In fact, as we will put in sharp relief in the next section, dynamics are not necessary to rule out large $c$ $\W_N$ CFTs with light operators. Using the $\W_N$ algebra alone, one can rule out light operators at large $c$ as follows.

The $\W_N$ algebra contains the sl($N$) algebra as its ``wedge subalgebra'', obtained by taking $c \to \infty$ and restricting to modes $W^{s}_m$ with $|m| < s$. This implies that we may compute the norms of light states, charged or uncharged, using only the sl($N$) algebra.
 In particular, let us consider the norm of $W^s_{-1}|h\ra$, which is a level-one, spin-$s$ descendant of an uncharged state $|h\ra$. Unitarity requires that $\la h|W^s_1W^s_{-1}|h\ra\geq 0$. We compute this using the known sl($N$) structure constants, which can be found in e.g. \cite{Gaberdiel:2011wb}. After some massaging,\footnote{The result requires just a single structure constant of sl($N$), equivalently, hs[$\lambda$] with $\l=N$. One uses the commutation relations and the primary condition to obtain
\e{}{\langle h|[W^s_1,W^s_{-1}]|h\rangle  = g^{ss}_{2s-2}(N) \langle h|L_0|h\rangle = g^{ss}_{2s-2}(N) h}
where $g^{ss}_{2s-2}(N)$ is a standard hs[$\l$] notation for structure constants. It can be computed using closed-form expressions given in \cite{Gaberdiel:2011wb}.} we find
\e{1.1}{\langle h|W^s_1W^s_{-1}|h\rangle = (-1)^s h\, \sigma^{2s-4} \prod_{k=2}^{s-1} (\lambda^2-k^2){s(s-1)\Gamma^2(s-1)\over (3/2)_{s-2}(5/2)_{s-2}} }
where $\sigma$ is a normalization constant.  Demanding that the spin-$s$ currents themselves have positive norm fixes $\sigma \in\mathbb{R}$ \cite{Gaberdiel:2012yb}. Given that and the manifest positivity of the product, the $(-1)^s h$ means that negative norm states exist at level one, even for $h>0$. 

\section{Unitarity Bounds at Finite $c$}\label{iii}

Unitarity of a 2d CFT implies that the theory must have a positive semi-definite  Kac matrix. In this section we use the $\mathcal{W}$-algebra commutation relations to compute the eigenvalues of the Kac matrix at level one, and impose positivity to derive constraints on the spectrum. The analysis applies only to the algebra $\W_N$, defined for example by Drinfeld-Sokolov reduction from $SU(N)$.  In this context, it is a significant extension of \cite{Perlmutter:2016pkf} and the results of the previous section, which required $c\gg 1$.

We perform this computation explicitly for theories with $\W_N$ symmetry for $N=3,4,5,6$. 
The commutation relations for these algebras appear in the literature, but with varying levels of accuracy.
The Jacobi relations for our implementation of the algebras have been verified using the Mathematica package {\tt OPEdefs.nb} by K. Thielemans \cite{Thielemans1991}.
Some of the Kac matrix calculations were performed with the help of the Mathematica package {\tt Virasoro.nb} by M. Headrick.\footnote{Available at \url{http://people.brandeis.edu/~headrick/Mathematica/}.}

Although the Kac matrix may be computed explicitly using the $\W_N$ algebra for small enough $N$, the computation becomes exceedingly tedious and intractable with increasing $N$. We will therefore turn to the Coulomb gas formalism in Sections \ref{cg1}--\ref{cg2} to extrapolate the results to $N>6$.

\subsection{Setup}
We will compute the analogue of the Kac matrix for the $\W_N$ theory, at level one. There are $N-1$ basis states given by
\begin{align}
	\ket{2} & = L_{-1} \ket{h,q_3,...,q_N}, \\
	\ket{3} &= W^{3}_{-1} \ket{h,q_3,...,q_N}, \\
	& \qquad \qquad \quad \vdots  \nonumber \\
	\ket{N} &= W^{N}_{-1} \ket{h,q_3,...,q_N}
\end{align}
The Kac matrix is
\begin{equation}
M_{ij} = \braket{i}{j}, \qquad i,j=2,...,N.
\end{equation}
The question is simply for what values of $h,q_i$ this matrix is positive semi-definite. 

The Kac determinant is known for $\W_N$ \cite{Mizoguchi:1989bw,Watts:1989bn}, so this may appear trivial, but in fact it is not. The closed form expression in the literature is in terms of auxiliary Coulomb gas parameters whose relation to the charges is nonlinear and unknown, so in particular, restricting to real $q_i$ (as required for unitarity) is not straightforward in that language. To overcome this obstacle, our strategy is to use explicit computations at low $N$ to guess the correct reality conditions, then to extend this to all $N$ using the Coulomb gas.

\subsection{Explicit calculations for $N=3,4,5,6$}\label{ss:explicit}

\subsubsection{$\mathcal{W}_3$}\label{sss:w3}

\begin{figure}[H]
	\centering
	\includegraphics[width=0.5\textwidth]{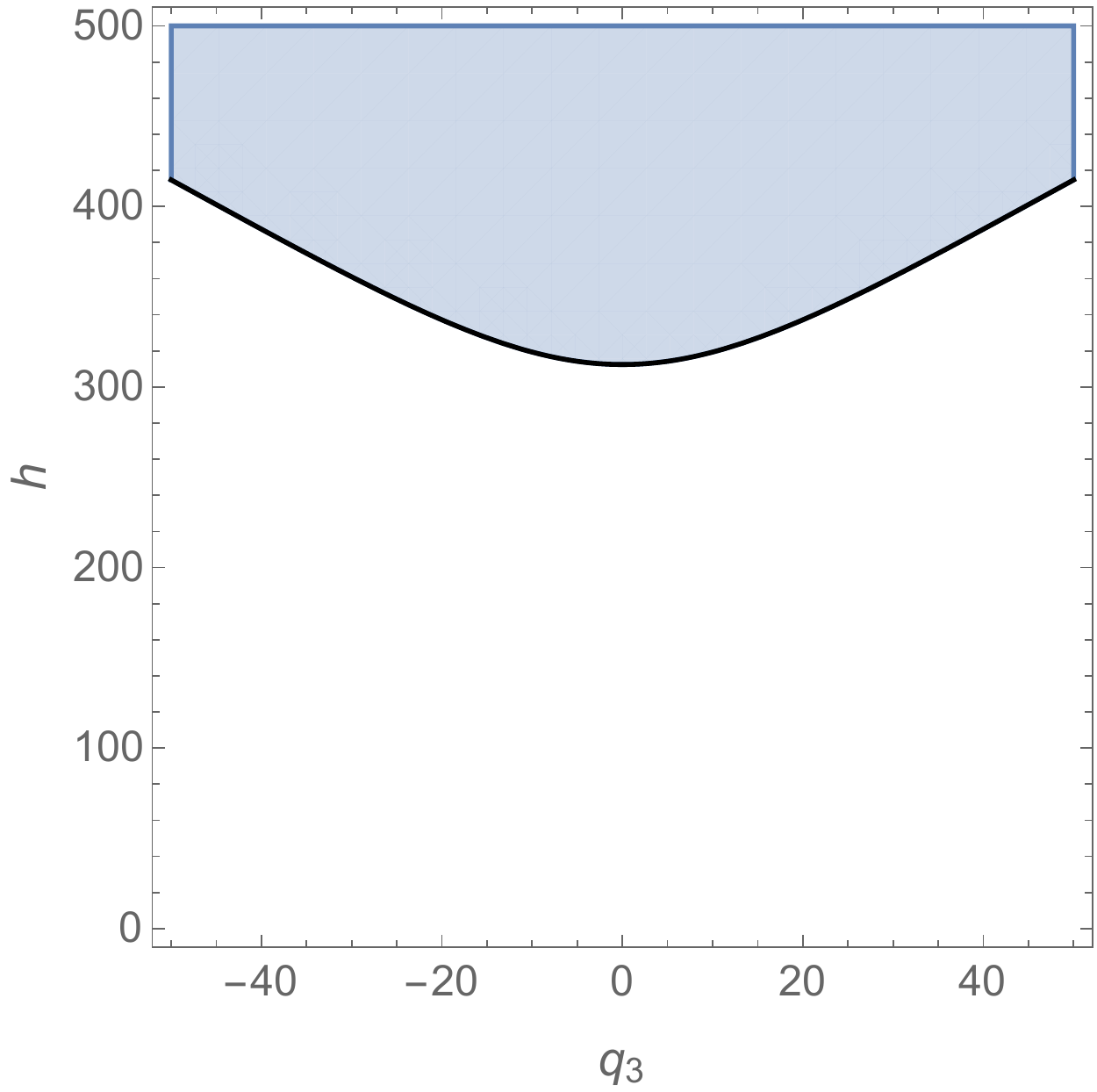}
	\caption{\small Exclusion plot for $\W_3$, with $c = 10^4$. The weight and charge $(h,q_3)$ of all primaries must fall in the shaded region, where $\lambda_2 > 0$.}\label{fig:W3Region}
\end{figure}

The $\mathcal{W}_3$ algebra can be found in e.g. \cite{Zamolodchikov1985,Bouwknegt:1992wg,Blumenhagen2009,Iles:2013jha,Mizoguchi:1988vk}.  The level-one Kac matrix is given by

\begin{equation}
M = \left( \begin{array}{cc}
\bra{h,q_3}L_1 L_{-1} \ket{h,q_3} &  \bra{h,q_3}L_1 W^{3}_{-1} \ket{h,q_3} \\
\bra{h,q_3}W^{3}_1 L_{-1} \ket{h,q_3} & \bra{h,q_3}W^{3}_1 W^{3}_{-1} \ket{h,q_3}
\end{array} \right),
\end{equation}
which can be explicitly computed as
\begin{equation}
M = \left( \begin{array}{cc}
2h & 3q_3 \\
3q_3 & \frac{h(2-c+32h)}{22+5c} 
\end{array} \right).
\label{det3}
\end{equation}
The eigenvalues are
\begin{align}\label{w3ev}
\lambda_1 &= \frac{h(46+9c+32h)+\sqrt{(42+11c-32h)^2 h^2 +36(22+5c)^2 q_3^2}}{2(22+5c)}\\
\lambda_2 &= \frac{h(46+9c+32h)-\sqrt{(42+11c-32h)^2 h^2 +36(22+5c)^2 q_3^2}}{2(22+5c)}\notag
\end{align}
We see that $\lambda_1$ is positive for all $h > 0$ and $c > 0$ but $\lambda_2$ can be negative. Primaries with $\lambda_2 < 0$ are excluded. In fact this bound has a simple global minimum: for arbitrary $q_3$, the conformal weight must satisfy
\be
h \geq \frac{c-2}{32} \ .
\ee
This is easily derived analytically from \eqref{w3ev}. Let us emphasize: this holds for all values of $c$. In the large-$c$ limit, this means that there can be no light primaries, of any charge, in a theory with $\W_3$ symmetry. This corroborates the causality analysis of Section \ref{s:causality}, though here the result is more general, because now we allow for $q_3 = 0$ and for a finite range of $h/c$. The full exclusion plot on the $(h,q_3)$ plane is shown in Figure \ref{fig:W3Region}. 
For nonzero $q_3$, one has a refined bound,
\e{}{{2h^2(32h-(c-2))\o 5c+22}\geq 9q_3^2}

\subsubsection{$\mathcal{W}_4$}

The $\mathcal{W}_4$ algebra can be found in \cite{Blumenhagen1990,Kausch:1990bn,Zhu1993}. Requiring the three eigenvalues of the level-one Kac matrix to be positive imposes complicated nonlinear constraints on $(h,q_3, q_4)$. The minimal conformal weight satisfying the constraints, for any values of the charges, is
\be\label{wfbb}
h \geq \frac{1}{30}(c-3) \ .
\ee
Again, the large-$c$ limit does not admit light states.  Note that in this case, the minimum is obtained for a state that has $q_3 = 0$ but $q_4 \neq 0$; the lower bound on uncharged states is stronger than \eqref{wfbb}.

\begin{figure}[H]
	\centering
	\includegraphics[width=0.5\textwidth]{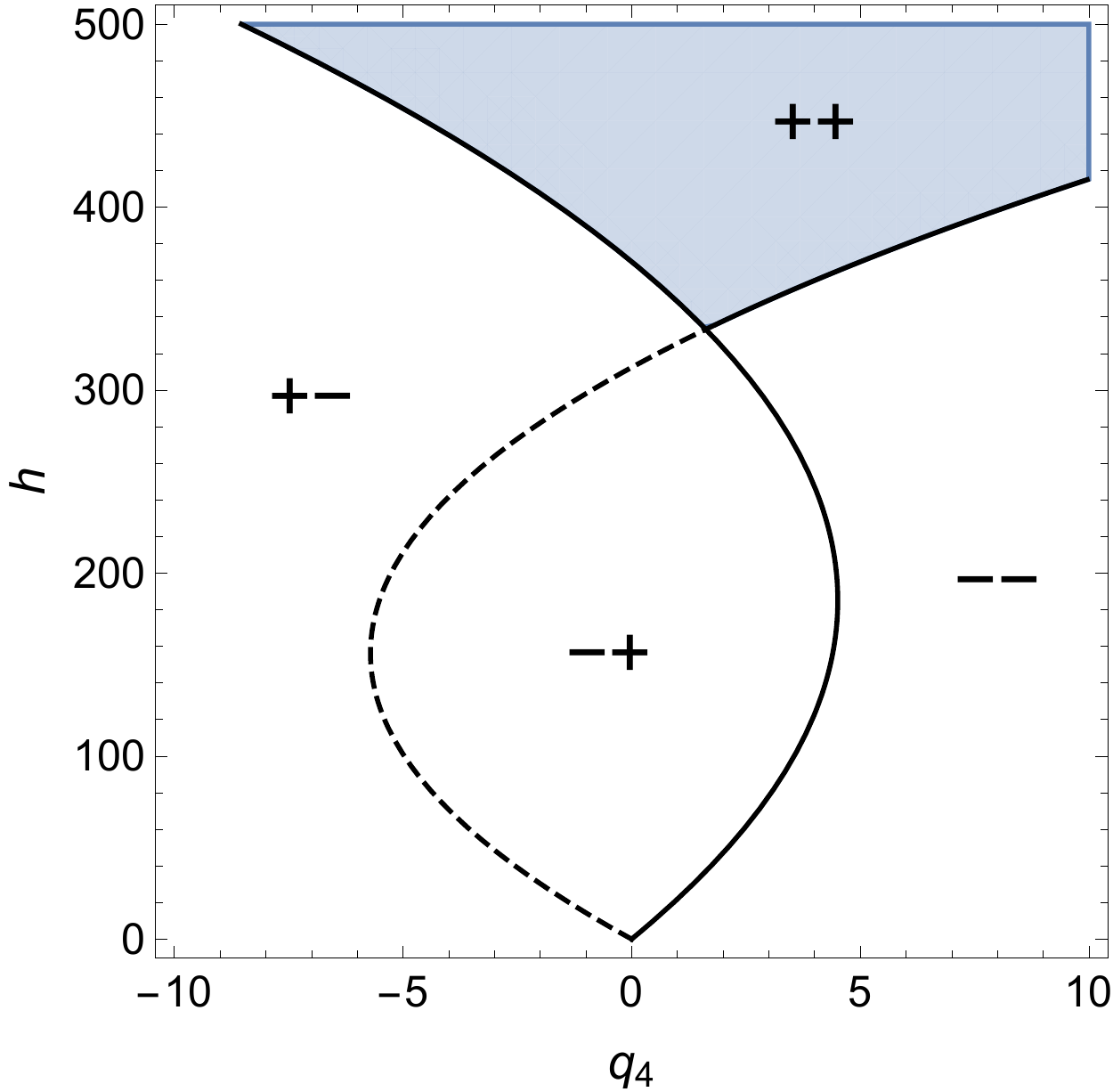}
	\caption{\small Exclusion plot for $\W_4$ charge $q_4$ and conformal weight $h$, with $c=10^4, q_3=0$. The shaded region is allowed, i.e. has all eigenvalues positive. One eigenvalue of the level-one Kac matrix is non-negative everywhere in the plotted region, and the pluses/minuses show the signs of the other two eigenvalues. The dashed curve shows the null states referred to in footnote \ref{footnote:null}.	\label{fig:W4Eigs}}
\end{figure}

We will not give a rigorous proof of the lower  bound \eqref{wfbb}, but it can be checked for any fixed value of $c$ as follows.  An $n\times n$ symmetric matrix is positive definite if and only if the $n$ leading principal minors are positive. That is, we impose $\det M_{(k)} > 0$ for $k=1\dots 3$, where $M_{(k)}$ is the matrix formed by the first $k$ rows and columns of $M$. Then we fix $c$ to some numerical value, and minimize $h$ subject to these polynomial constraints using Mathematica. The result is \eqref{wfbb}. We have checked this for thousands of random rational values of $c>3$ and for $c = \infty$ (with all charges scaling as $c$).\footnote{A subtlety here is that unitarity of the level-one Kac matrix only requires that $M$ is positive \textit{semi-}definite, while the algorithm just describes actually finds inf $h$ subject to the constraint that $M$ is positive definite, with strict inequalities.  These bounds are in fact different: the minimal $h$ with non-negative eigenvalues is zero. This is because there is a sublocus in charge space, with null states at level one, which extends all the way down to $h=0$. In Figure \ref{fig:W4Eigs}, this sublocus is the dashed curve.  We are excluding these degenerate cases from the discussion.  For $\W_4$, at least for $q_3 = 0$, the level-2 Kac matrix imposes new constraints and restricts even these null states to $h \geq (c-3)/40$, so in this case the conclusion that there are no light states at large $c$ applies also to degenerate representations. For $N\geq 5$ we will not consider the degenerate case.\label{footnote:null}}

An exclusion plot is shown in Figure \ref{fig:W4Eigs}.

\subsubsection{$\mathcal{W}_5$}

The $\W_5$ algebra is derived in \cite{Hornfeck1992,Zhu1993}. From the level-one Kac matrix, we claim the global minimum conformal weight is
\begin{equation}\label{w4b}
h \geq \frac{3}{80}(c-4).
\end{equation}
The bound is obtained for $q_3 = q_5 = 0$, but $q_4 \neq 0$. Again we do not have a proof, but the evidence for this lower bound is as follows. First, we have checked numerically that it is a local minimum for various values of the parameters. Second, if we assume the odd charges vanish, then we have checked that \eqref{w4b} is a global minimum for thousands of random values of $c$, using exactly the same method described for $\W_4$ above. Third, without assuming the odd charges vanish, we have scanned numerically over the charge space with $c = 4.4, 5.6, 8, 15, 20, 40, 80, 200$, and the minimum \eqref{w4b} is the lowest value of $h$ obtained in these scans.  The numerical scan uses a simple grid over a finite region in charge space, so it is not guaranteed to find a global minimum, but we have scanned over a region approximately 100 times larger than the values of $h, q_4$ at the minimum, without finding any violation of \eqref{w4b}.

\subsubsection{$\mathcal{W}_6$}

By the same methods, the $\W_6$ algebra \cite{Hornfeck1992} leads to the lower bound
\begin{equation}\label{w6b}
h \geq \frac{4}{105}(c-5).
\end{equation}
In this case the numerics are more difficult, and while we believe this to be a global lower bound, the evidence is not overwhelming. We have confirmed numerically that it is a local minimum, and  by scanning over charge space at various values of $c$ we failed to find any lower global minimum. However the scan is now limited by computational time to a relatively small region around the point saturating \eqref{w6b}. 

\subsection{Coulomb gas review}\label{cg1}

The explicit results for $N=3,4,5,6$ match the result \eqref{result} quoted in the introduction. Now we aim to extend this to all $N$.   The strategy is to identify the null states at the global minimum for $N \leq 6$, find a pattern, then extrapolate.

The $\W_N$ algebras can be realized by $N-1$ free fields. Closely following the review \cite{Bilal:1991eu}, our conventions are as follows.
For $SU(N)$, we have $N-1$ simple roots which are normalized such that the Cartan matrix, $e_i\cdot e_j$ is $2$ along the diagonal. They can be represented as vectors in $\mathbb{R}^N$, as $e_1 = (-1,1,0,\dots)$, $e_2  = (0,-1,1,0,\dots)$, etc. The $N-1$ fundamental weights $\lambda_i$ are dual to the simple roots:
\begin{align}
	\lambda_i\cdot e_j=\delta_{ij}.
\end{align}
We also define the weights of the vector representation $h_\mu$ with $\mu=1,...,N$:
\begin{align}
	h_\mu=\lambda_1-\sum_{i=1}^{\mu-1}e_i.
\end{align}
The Weyl vector is given by $\rho=-\sum_i^{n}\mu h_\mu$. 

In the Coulomb gas construction, the stress tensor  $T$ and the higher spin fields $W^{s}$ are constructed from free fields $\phi_i$ where the index $i=1,...,N-1$. A convenient choice of basis in this $(N-1)$-dimensional space is provided by the fundamental weights of $SU(N)$. The holomorphic stress tensor can be derived from the action 
\begin{align}
	S=\frac{1}{8\pi}\int d^2z \sqrt{g}(\partial\phi\cdot\partial\phi+2i\alpha_0R\rho\cdot\phi)
\end{align} 
and is given by
\begin{align}
	T(z)=-\frac{1}{2}:\!\partial\phi(z) \cdot \partial\phi(z)\!:+2 i \alpha_0 \rho\cdot\partial^2\phi(z),
\end{align}
where $\rho$ is the Weyl vector and $\alpha_0$ is a constant. The corresponding central charge is related to $\a_0$ as
\begin{align}
%	c&=(N-1)(1- 4\alpha_0^2 N(N-1)),\\
	\alpha_0&=\pm \frac{\sqrt{-c+N-1}}{2 \sqrt{N \left(N^2-1\right)}}.
\end{align}

The vertex operators defined as
\e{}{V_{\Lambda}\,\equiv\,\, :\!\exp(i \Lambda\cdot\phi(z))\!:}
are Virasoro primaries with holomorphic weight 
\begin{align}
	\Delta_2(\Lambda)=\frac{1}{2}\Lambda\cdot(\Lambda-4\alpha_0 \rho).\label{weightform}
\end{align}
The highest weight states of the $\mathcal{W}$-algebra are obtained from the action of the vertex operators on the $SL(2,\mathbb{C})$ invariant vacuum,
\begin{align}
\ket{\Lambda} \equiv V_\Lambda(0).
\end{align}
These states can be labelled by their eigenvalues under the zero modes of the stress-tensor and the higher spin currents:
\begin{align}
	L_0\ket{\Lambda}=\Delta_2\ket{\Lambda},\hspace{.5in}
	W_0^{s}\ket{\Lambda}=\Delta_s\ket{\Lambda}.
\end{align}
To construct the higher spin fields $W^{s}$, consider the currents $u_s$ defined by the so called quantum Miura transformation:
\begin{align}
	:\prod_{\mu=1}^N[ 2i\alpha_0\partial+h_\mu\cdot\partial\phi(z)]:\,=(2i\alpha_0)^n\partial^N+\sum^n_{s=1}(2 i \alpha_0)^{n-s}u_s(z)\partial^{N-s},
\end{align}
where $h_\mu$ are the weights of the vector representation of $SU(N)$ as defined above. These generate the $\W$ symmetry, but they are not Virasoro primaries (see e.g.  \cite{Bouwknegt:1992wg}), as can be confirmed by computing the OPE with the stress tensor. Virasoro primaries can be constructed from combinations of the normal ordered products of these currents as well as their derivatives. This implies that the eigenvalues of the highest weight states with respect to the zero modes of the higher spin charges can be constructed from the eigenvalues of the zero modes of the currents $u_s$. For example
\begin{align}
	W^{3}&=u_3-\frac{n-2}{2}2i \alpha_0\partial u_2,\notag\\
	\Delta_3(\Lambda)&=\tilde{\Delta}_3(\Lambda)+2 i \alpha_0(n-2) \tilde{\Delta}_2(\Lambda)
\end{align}
where $u_s|\Lambda\ra = \tilde{\Delta}_s(\Lambda)|\Lambda\ra$. The $\tilde\D_s(\Lambda)$ can be written in closed form:
\begin{align}
	\tilde{\Delta}_s(\Lambda)=(-i)^s\sum_{\mu_1>\mu_2>...\mu_s}\prod_{n=1}^s[h_{\mu{_n}}\cdot\Lambda+2\alpha_0(s-n)].
\end{align}
Finally, we need to understand when a primary $|\Lambda\rangle$ has null descendants.  Let $\alpha_{\pm}$ be the two solutions of
\begin{align}
	\alpha_{\pm}^2-2\alpha_0\alpha_{\pm}=1.
\end{align}
Null descendants appear when the vector $\Lambda$, expanded in the basis of fundamental weights, takes the form 
\begin{align}
	\Lambda&=\sum_{j=1}^{N-1}((1-l_j)\alpha_-+(1-l_j')\alpha_+)\lambda_j
	\label{nullvec1}
\end{align}
with at least one of the $\ell_i$'s equal to a positive integer (see e.g. \cite{Bilal:1991eu} for the derivation). In what follows it will be convenient to parameterize $\Lambda$ as
\be\label{nullvec2}
\Lambda = \sum_{j=1}^{N-1}i L_j \lambda_j \ .
\ee
If any $L_i$ is zero, then there is a null state at level one.

There is a closed-form expression for the Kac determinant with $\W_N$ symmetry \cite{Mizoguchi:1989bw,Watts:1989bn}. At level one, 
\begin{align}
	\text{det} M\propto \left(\prod_{e\in \Delta} (\Lambda-2 \alpha_0 \rho)\cdot e-(\alpha_++\alpha_-)\right),
\end{align}
where $\Delta$ is the set of adjoint weight vectors of $SU(N)$. Evaluating this expression for the case of $\W_3$ yields
\begin{align}
	\det M=-\frac{L_1 L_2 \left(\sqrt{6} \sqrt{2-c}-6 i L_1\right) \left(\sqrt{6} \sqrt{2-c}-4 i L_1-4 i L_2\right)}{108 (5 c+22)}\notag\\
	\times\left(\sqrt{6} \sqrt{2-c}-6 i L_2\right) \left(\sqrt{6} \sqrt{2-c}-12 i L_1-12 i L_2\right).
\end{align}
This can be easily verified to match determinant of \eqref{det3} using 
\begin{align}
	h&=\Delta_2(\Lambda),\notag\\
	q_3&=-\frac{4 i \sqrt{3}}{\sqrt{5 c+22}} \Delta_3(\Lambda).
\end{align}
Furthermore, one can verify that setting any of the $L_i$ to zero yields a vanishing determinant for any $N$, consistent with the description of the null states provided above.

\subsection{Unitarity bound for general $N$}\label{cg2}

To recap, in the Coulomb gas formalism, states are labeled by the parameters $(L_1, \dots L_{N-1})$ defined in \eqref{nullvec2}. In principle, our task is to find the range of $L_i$'s allowed by unitarity, then evaluate the minimal weight using \eqref{weightform}. This appears to be difficult in general, but we can make some progress guided by the fact that this minimum will always appear at the boundary of the allowed region, so it will lie at point with one or more null states. For example, it is apparent from Figure \ref{fig:W3Region} that the $\W_3$ optimum has one null state, while the optimum for $\W_4$ sits at the intersection of two curves in Figure \ref{fig:W4Eigs} so has two null states.  We will use these low-lying examples to guess a pattern in the null states at the optimum, and conjecture a lower bound on the conformal weight for all $N$.

\subsubsection{$\mathcal{W}_3$ Example}
The $\mathcal{W}_3$ Kac matrix at level one is given by \eqref{det3} and the eigenvalues are given by \eqref{w3ev}. Setting $\lambda_2=0$  we find the curve $h_{min}(q_3)$ corresponding to null states as shown by the black curve in Figure \ref{fig:W3Region}.

This curve may also be obtained starting from the null state equation \eqref{nullvec1}. For generic $\Lambda=\sum i L_j \lambda_j$ with $L_j\in \mathbb{R}$ we find
\begin{align}\label{wtwa}
	\Delta_2&=\frac{1}{12} \left(\sqrt{6} \sqrt{c-2} L_1+\sqrt{6} \sqrt{c-2} L_2-4 \left(L_1^2+L_2 L_1+L_2^2\right)\right)\\
	\Delta_3&=\frac{1}{216} (L_1-L_2) \left(-6 \sqrt{6} \sqrt{c-2} L_1-6 \sqrt{6} \sqrt{c-2} L_2+3 c+8 (2 L_1+L_2) (L_1+2 L_2)-6\right).\label{wtwb}
\end{align}
As described above, setting one of the $L_i$'s to zero gives a null state at level one. If we choose $L_2 = 0$ and use $\sqrt{\frac{22+5c}{48}}i q_3=\Delta_3$, the two equations \eqref{wtwa},\eqref{wtwb} reduce to exactly the same equation $h = h_{min}(q_3)$ found in section \ref{sss:w3}. The constant of proportionality relating $q_3$ to $\Delta_3$ follows from the conventions chosen to normalize the $\mathcal{W}_3$ current. 

To find the optimum, in the Coulomb gas language, we then extremize $\Delta_2(L_1)$ with respect to $L_1$.  This reproduces the same global lower bound as before, $h \geq \frac{c-2}{32}$. Note that this extremum is not necessarily a minimum for real $L_1$.

\subsubsection{General case}

Motivated by this success we will proceed for general $N$ in the same way: set some of the $L_i$'s to zero, then extremize $\Delta_2$ over the remaining $L_i$. This procedure guarantees that we will find null states but does not necessarily guarantee that we will find the global minimum $h$. To find the global minimum requires that we pick the correct null state equations and choose the correct extremum. Working through the low-level examples as we did for $\W_3$ reveals a simple pattern that we conjecture extends to all $N$.

The details depend on whether $N$ is even or odd. For $N$ even, we take the ansatz that the odd $L_i$'s vanish,
\be
L_1 = L_3 = \cdots = L_{N-1} = 0
\ee
then extremize the weight
\begin{align}
	\Delta_2(\Lambda)&=\frac{1}{2}\Lambda\cdot (\Lambda-4\alpha_0 \rho)\notag
\end{align}
with respected to the remaining, even $L_i$'s:
\begin{align}
	\frac{\partial \Delta_2(\Lambda)}{\partial L_a}&=
	-\sum_{b \mbox{\ \footnotesize even}}L_b \lambda_b\cdot\lambda_a-2i\alpha_0\lambda_a\cdot\rho\notag\\
	&=-\frac{1}{N}\sum_{b \mbox{\ \footnotesize even}}[a(N-b)\delta_{a\leq b}+b(N-a)\delta_{a>b}]L_b-i \alpha_0 a(N-a)\notag\\
	&=0, \hspace{1.4in}  \qquad (a \mbox{\ even}).\notag
\end{align}
This can be solved by setting all of the even $L_i$'s to $L_i = -4 i \alpha_0$. So in the end we find the state
\be\label{soleven}
N \ \mbox{even}: \qquad \Lambda_* = 4\alpha_0 \sum_{i \mbox{\ \footnotesize even}}\lambda_i
\ee

In the case where $N$ is odd we instead set the even-indexed $L_i$ to zero, and extremize $\Delta_2$ with respect to the odd $L_i$. We find the extremum
\be\label{solodd}
N \ \mbox{odd}:  \qquad \Lambda_* = 3\alpha_0 \lambda_1 + 4 \alpha_0 \sum_{i>1 \mbox{\ \footnotesize odd}}\lambda_i \ .
\ee
Evaluating the weight of these states gives, for all $N$,
\be
h_{crit} \equiv \Delta_2(\Lambda_*) = {c-(N-1)\over 24}\left(1-  \frac{6\floor{\frac{N}{2}}}{N(N^2-1)}\right)
\ee
This agrees with the explicit global minima for $N=3,4,5,6$ constructed in section \ref{ss:explicit}, so at least in those cases, we have found the Coulomb gas representation of these minimal weight states. Furthermore, by construction the states \eqref{soleven}, \eqref{solodd} have null descendants at level one, and extremize $h$ with respect to the remaining parameters. We take this as evidence that we have indeed identified the minimal states for all $N$, leading to the lower bound $h \geq h_{crit}$ discussed in the introduction.

\section{A Modular Bootstrap for $\W_3$ CFTs}

In this section we explore the constraints of modular invariance for ${\cal W}_N$ CFTs.  
In addition to modular invariance of the torus partition function, we will also 
use the modular transformation properties of the partition function with insertions of higher spin charges.  
The result will be a set of upper bounds on the dimension of the lightest non-identity primary operator in the theory.
These constraints are in tension with our earlier bounds arising from the unitarity of the Kac matrix, which ruled out primary states which were too light.  We will not, however, obtain a direct contradiction which would rule out ${\cal W}_N$ theories with $c>N-1$.
We will focus on the ${\cal W}_3$ case, where the modular transformation properties of characters with insertions of higher spin currents were worked out in \cite{Iles:2014gra}.  
It is in principle straightforward (although in practice tedious) to generalize this to higher $N$.  
A similar modular bootstrap strategy for ${\cal W}_3$ CFTs was discussed in \cite{Apolo:2017xip}.

We will follow the modular bootstrap approach initiated in \cite{Hellerman:2009bu}, where modular invariance of the torus partition was used to place an upper bound on the dimension of the lightest non-identity primary operator. This bound is simplest to state at large central charge:
\begin{equation}\label{bilbo}
h_1 + {\bar h}_1 \le {c\over 6} +{\cal O}(1) 
\end{equation}
where ${\cal O}(1)$ denotes a numerical constant which remains finite as $c\to\infty$. 
This result assumes only unitarity, modular invariance and that the theory has a discrete spectrum with a normalizable ground state.  

\ssec{Review of modular bootstrap for the partition function}
Let us start by recalling the derivation of these bounds in the Virasoro case.  We begin by writing the torus partition function as a sum of two terms: 
\begin{equation}\label{frodo}
Z(\tau, {\bar \tau}) \equiv \Tr (q^{L_0-{c\o 24}} {\bar q}^{{\bar L}_0-{c\o 24}}) = 
\Tr_{\vac} (q^{L_0-{c\o 24}} {\bar q}^{{\bar L}_0-{c\o 24}}) + 
\sum_p  
\Tr_{V_p} (q^{L_0-{c\o 24}} {\bar q}^{{\bar L}_0-{c\o 24}})~.
\end{equation}
where $ q=e^{2\pi i \tau}$. The first term is the contribution of the vacuum state and its descendants, and the second is the contribution from all of the non-identity primaries, labelled by an index $p$.  Here $V_p$ is the Verma module built on the primary state of weight $(h_p, {\bar h}_p)$; we assume that $c$ is large enough that there are no null states.  
This allows us to write the condition of modular invariance, $Z(\tau, {\bar \tau}) = Z(-1/\tau, -1/{\bar \tau})$, as 
\es{frodo}{&\left(\Tr_{\vac} (q^{L_0-{c\o 24}} {\bar q}^{{\bar L}_0-{c\o 24}}) - \Tr_{\vac} ({\hat q}^{L_0-{c\o 24}} \hat{\bar q}^{{\bar L}_0-{c\o 24}})\right)\\
+ 
\sum_p
&\left(
\Tr_{V_p} (q^{L_0-{c\o 24}} {\bar q}^{{\bar L}_0-{c\o 24}}) - \Tr_{V_p} ({\hat q}^{L_0-{c\o 24}} \hat{\bar q}^{{\bar L}_0-{c\o 24}})
\right) = 0 }
where ${\hat q} = e^{-2\pi i / \tau}$. 
The traces appearing in this equation are the usual characters of the Virasoro algebra.
We then proceed by evaluating this expression order by order in a derivative expansion around $\tau=i$,  where $q={\hat q}$.  This leads to an expression of the form
\begin{equation}\label{samwise}
\Xi_{\vac}^{(n, {\bar n})}(c) + \sum_p \Xi^{(n, {\bar n})}(c,h_p, {\bar h}_p) =  0
\end{equation}
where 
\begin{equation}
\Xi_{\vac}^{(n, {\bar n})} \equiv 
(q \partial_q)^n ({\bar q} \partial_{\bar q})^{\bar n} 
\left. \left(\Tr_{\vac} (q^{L_0-{c\o 24}} {\bar q}^{{\bar L}_0-{c\o 24}}) - \Tr_{\vac} ({\hat q}^{L_0-{c\o 24}} \hat{\bar q}^{{\bar L}_0-{c\o 24}})\right)
\right|_{\tau=i}
\end{equation}
with a similar expression for $\Xi^{(n, {\bar n})}$.  
For small $(n, {\bar n})$ these constraints can be studied analytically, where they lead to (\ref{bilbo}) \cite{Hellerman:2009bu}.  At higher $(n, {\bar n})$ they can be studied numerically \cite{Friedan:2013cba, Collier:2016cls}.  

The computation described generalizes immediately to ${\cal W}_N$ theories.  One starts by organizing the partition function into representations of ${\cal W}_N$, and replacing (\ref{frodo}) by a sum over ${\cal W}_N$ primaries rather than Virasoro primaries.  
The vacuum and Verma module characters are $|\chi_{\vac}(\t)|^2$ and $|\chi_p(\t)|^2$, with
\e{mordor}{\chi_p(\tau) =  \frac{q^{h_p-\frac{c-N+1}{24}}}{\eta(q)^{N-1}}~, \quad \chi_{\vac}(\tau)  = \chi_0(\t)\prod_{n=1}^{N-1} (1-q^n)^{N-n}~.}
Just as in the Virasoro case, it is straightforward to work out the modular transformations of these characters.
The result is that one can write the equations (\ref{samwise}) explicitly in any ${\cal W}_N$ theory.
The modular bootstrap analysis can then be carried out, just as in the Virasoro case.  
In fact, the difference between the ${\cal W}_N$ and Virasoro computations is unimportant at large $c$.  In particular, at large $c$ one recovers precisely the bound (\ref{bilbo}) \cite{Hartman:2014oaa,Qualls:2015bta}.
The only important difference is that (\ref{bilbo}) is now interpreted as a bound on the dimension of the lightest ${\cal W}_N$-primary, rather than the lightest Virasoro primary.

\ssec{Spin-3 charged modular bootstrap}
Instead, we will describe a somewhat more powerful technique.
We will consider the modular transformation properties of the torus partition function with insertions of the higher-spin charge $W_0$.  We will focus on the ${\cal W}_3$ case, where Iles and Watts \cite{Iles:2014gra} have already determined the relevant modular transformation properties.  These authors derived the elegant result
\begin{align}
\label{modtrans}
\Tr\left(W_0^2\hat{q}^{L_0-\frac{c}{24}}\hat{\bar{q}}^{L_0-\frac{c}{24}}\right)=&\tau^6 \Tr\left(W_0^2q^{L_0-\frac{c}{24}}{\bar{q}}^{{\bar L}_0-\frac{c}{24}}\right)\notag\\
&+\frac{\alpha}{i \pi \tau}\left[D^{(2)}D^{(0)}+\frac{c}{1440}E_4(q)\right]\Tr\left(q^{L_0-\frac{c}{24}}\bar{q}^{{\bar L}_0-\frac{c}{24}}\right)
\end{align}
where $\alpha=\frac{16}{22+5c}$. 
Here $E_n(q)$ is an Eisenstein series and 
\begin{equation}
D^{(r)}\equiv q \partial_q - {r\over 12} E_2
\end{equation}
is the Serre derivative.  
To understand this formula, note that the
the zero mode of any dimension-$6$ primary operator will transform like a modular form of weight 6; this transformation rule is the first line of the equation.  
The second line comes from the fact that, while $W_0^2$ is not the zero mode of a dimension 6 operator, it is 
 -- using the ${\cal W}_3$ algebra -- possible to find a primary operator whose zero mode is a linear combination of 
 $W_0^2$ and a polynomial in $L_0$.  The second line of this formula reflects the ``anomalous" modular transformation properties coming from the addition of this polynomial in $L_0$. 
 
The important point is that, just like the partition function, the coefficients in the $q$-expansion of  $\Tr\left(W_0^2\hat{q}^{L_0-\frac{c}{24}}\hat{\bar{q}}^{{\bar L}_0-\frac{c}{24}}\right)$
are positive.  
Thus one can apply the same modular bootstrap logic to this expression.
In particular, we separate out the vacuum contribution to write 
\eqref{modtrans} as
\begin{align}
0=&
\Big(\text{Tr}_{\vac}
\left(W_0^2\hat{q}^{L_0-\frac{c}{24}} \right) \bar{\chi}_{\vac}(\hat{\bar{q}})
-\tau^6 
\text{Tr}_{\vac}
\left(W_0^2 q^{L_0-\frac{c}{24}}\right) 
{\bar \chi}_{\vac} ({\bar q})
\notag\\
&
\hspace{.5in}
-
\frac{\alpha}{i \pi\tau}
\left[D^{(2)}D^{(0)}+\frac{c}{1440}E_4(q)\right]
{\chi}_{{\vac}}({q})
\bar{\chi}_{{\vac}}({\bar q})
\Big)
\notag\\
&
+
\sum_p \Big(\text{Tr}_{V_p}
\left(W_0^2\hat{q}^{L_0-\frac{c}{24}} \right) 
\bar{\chi}_{p}(\hat{\bar{q}})
-\tau^6 
\text{Tr}_{V_p}
\left(W_0^2 q^{L_0-\frac{c}{24}}\right) 
{\bar \chi}_{p} ({\bar q})
\notag\\
&
\hspace{.5in}
-
\frac{\alpha}{i \pi\tau}
\left[D^{(2)}D^{(0)}+\frac{c}{1440}E_4(q)\right]
{\chi}_{p}({q})
\bar{\chi}_{p}({\bar q})
\Big)
\label{voldemort}
\end{align}
Here $\chi_{\vac}$ and $\chi_p$
are the characters of the ${\cal W}_3$ algebra \eqref{mordor};
it is then straightforward to compute the derivatives $D^{(2)} D^{(0)}$ of these characters appearing in \eqref{voldemort}. 
The Verma module trace with the insertion of $W_0^2$ was computed in \cite{Iles:2013jha}:
\begin{align}
\text{Tr}_{V_p}\left(W_0^2q^{L_0-\frac{c}{24}}\right)=&\left(
\frac{1}{864}\alpha c-24 h_p-2\left(E_2(q)^2-E_4(q)\right)+\frac{\alpha c+30(E_2(q)E_4(q)-E_6(q))}{8640}\right.\notag\\
&\left.+\frac{1}{648}\alpha \left(E_2(q)^3-3E_4(q)E_2(q)+2E_6(q)\right)+w_p^2\right)\frac{q^{h_p-\frac{c-2}{24}}}{\eta(q)^2},\notag\\
\label{vtrace}
\end{align}
The important feature of this expression is that it depends on both the dimension $h_p$ and the ${\cal W}_3$-charge $w_p$ of the primary state -- this means that, unlike the Virasoro case, our final bound will involve the charge $w_p$ of the lightest non-trivial primary operator. 
We can now use \eqref{vtrace}, along with equation (2.14) of \cite{Iles:2013jha} (derived using the Kazhdan-Lusztig algorithm \cite{DeVos:1995an}), to compute the trace in the vacuum module: 
%\begin{align}
\es{}{\text{Tr}_{\vac}\left(W_0^2q^{L_0-\frac{c}{24}}\right)&=
\text{Tr}_{V_{0,0}}\left(W_0^2q^{L_0-\frac{c}{24}}\right)-\text{Tr}_{V_{1,w_-}}\left(W_0^2q^{L_0-\frac{c}{24}}\right)-\text{Tr}_{V_{1,-w_-}}\left(W_0^2q^{L_0-\frac{c}{24}}\right)\\
&+\text{Tr}_{V_{3,w_+}}\left(W_0^2q^{L_0-\frac{c}{24}}\right)+\text{Tr}_{V_{3,-w_+}}\left(W_0^2q^{L_0-\frac{c}{24}}\right)-\text{Tr}_{V_{4,0}}\left(W_0^2q^{L_0-\frac{c}{24}}\right)}
%\end{align}
where $V_{h,w}$ is the Verma module built on a primary with dimension $h$ and charge $w$, and 
\begin{equation}
w_{\pm}=\frac{\sqrt{6} (\pm t+1)}{\sqrt{(5-3 t) (5 t-3)}},~~~~~ t=\frac{1}{48} \left(\sqrt{c^2-100 c+196}-c+50\right)~.
\end{equation}
Finally, one can use the modular transformation properties of $\eta(q)$ and $E_{n}(q)$ to write the various terms in \eqref{voldemort} evaluated at $\hat{q}$ in terms of these functions evaluated at $q$, rather than $\hat{q}$.  
The result is a final expression for
\eqref{voldemort}, which we write as 
\begin{equation}
\Xi_{\vac}(\tau,c)+\sum_p\Xi(\tau,c,h_p,w_p)=0
\label{schematic}
\end{equation}
where $\Xi_{\vac}$ and $\Xi$ can be written in terms of elementary modular functions.  We will not write the explicit expressions here, as they are lengthy and
not particularly illuminating, instead leaving them to the reader's imagination.

We now proceed as in the Virasoro case, extracting constraints on the spectrum of theory by acting with linear functionals on \eqref{schematic}.  
In particular, we expand order by order in derivatives around $\tau=i$ to obtain
\begin{align}
\Xi^{(n)}_{\vac}(c)+\sum_p \Xi^{(n)}(c,h_p,w_p)=0
\label{bsexpansion}
\end{align}
where $\Xi^{(n)} \equiv (q\partial_q)^n\Xi|_{\tau=i}$.
These equations give a set of linear constraints on the spectrum, which can be analysed using the standard techniques of linear programming.
In particular, if one can find a set of real numbers $\{\alpha_k\}$, $k=0,\dots,N_{max}$, such that
\begin{align}
\sum_{k=0}^{N_{\text{max}}}\alpha_k \Xi^{(k)}_{\tiny{\text{vac}}}(c)>0 
\label{vacuumpos}
\end{align}
then equation \eqref{bsexpansion} implies the existence of an operator such that
\begin{align}
\sum_{k=0}^{N_{\text{max}}}\alpha_k \Xi^{(n)}(c,h,w)<0. 
\label{nonvacneg}
\end{align}
These constraints can be analyzed numerically to find optimal bounds for any fixed $c$.  We have performed this numerical analysis up to sixth order in derivatives.  We have also performed an analytic computation to obtain bounds which are valid at large $c$.  Our numerical results are qualitatively similar to this analytic result, so we will just describe our analytic result at large $c$.

We begin by choosing a particular linear combination of the constraints \eqref{bsexpansion} which annhilate the vacuum contribution.
In particular, we define
\begin{align}
\Psi^{(i,j)}(c,h,w)= \left(\frac{\Xi^{(j)}_{\vac}(c)}{\Xi^{(i)}_{\vac}(c)}\right) \Xi^{(i)}(c,h,w)-\Xi^{(j)}(c,h,w)
\end{align}
so that \eqref{bsexpansion} implies that for any $i,j$ we must have
\begin{align}
\sum_p \Psi^{(i,j)}(c,h_p,w_p)=0~.
\end{align} 
This implies that there must at least one operator with $\Psi^{(i,j)}>0$, and at least one operator with $\Psi^{(i,j)}<0$.  This leads to simple analytic bounds on the spectrum at large $c$.
For example, at large $c$ we have
\begin{align}
\Psi^{(1,3)}=\frac{\sqrt{\pi } c^4 \left((\delta -1) \delta  \left((2 \delta -1)^3+135 \eta ^2\right)+15 \eta ^2\right)}{155520 \Gamma \left(\frac{1}{4}\right)^2}+\mathcal{O}(c)^3
\label{largecbnd}
\end{align}
where we have defined the rescaled charges
\begin{align}
\delta\equiv \frac{12h}{c},\hspace{1in}\eta\equiv \frac{12 w}{c}.
\end{align}
\begin{figure}
	\centering
	\includegraphics[scale=.6]{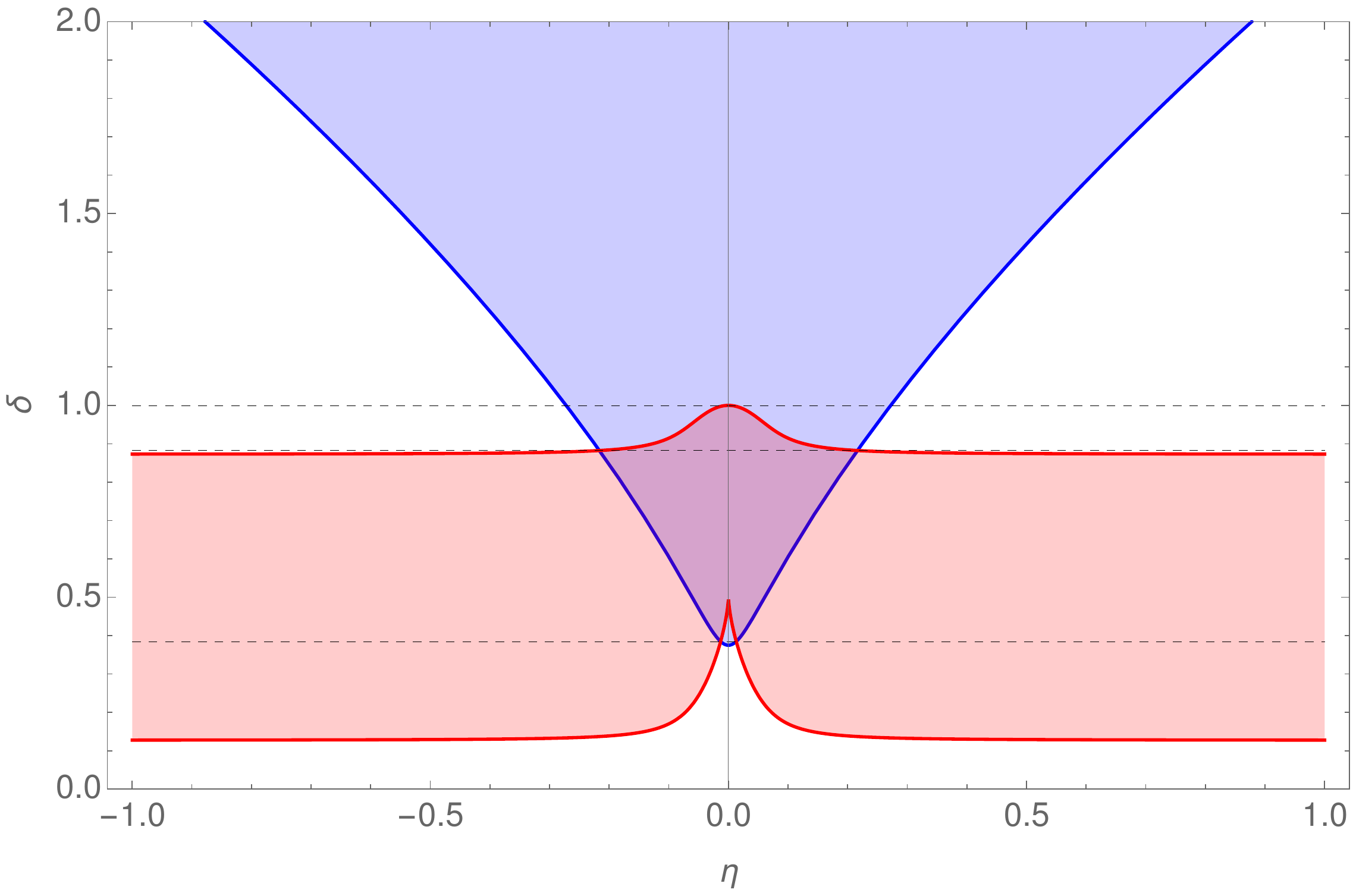}
	\caption{\label{fig1} The red region is where $\Psi^{(1,3)}>0$; modular invariance implies the existence of an operator in this region. The blue region is the unitarity bound set by the positivity of the Kac matrix. }	
\end{figure}
The sign of this function is plotted in Figure \ref{fig1}.\footnote{Figure \ref{fig1} shows the positive region for the leading ${\cal O}(c^4)$ part of $\Psi^{(1,3)}$.  The plot does not change qualitatively when one studies $\Psi^{(1,3)}$ at large but finite $c$.}
There must be at least one state in the red region, where $\Psi^{(1,3)}$ is positive. 
The boundary of this region (where $\Psi^{(1,3)}=0$) is given by the curves
\begin{align}
\eta(\delta)=\pm\frac{i \sqrt{\delta -1} \sqrt{\delta } (2 \delta -1)^{3/2}}{\sqrt{135 \delta ^2-135 \delta +15}} ~.
\end{align}
For uncharged operators, this coincides with the usual bound \eqref{bilbo} (i.e. $\delta<1$). For charged operators, this bound is stronger; our modular bound intersects the Kac matrix bound at $\delta\approx 0.882$. 

These bounds can be improved by considering linear combinations of the form $\sum_{i,j} \gamma_{i,j} \Psi^{(i,j)}$.
For any choice of constants $\gamma_{i,j}$, there must be an operator with $\sum_{i,j} \gamma_{i,j} \Psi^{(i,j)}<0$ and also an operator with $\sum_{i,j} \gamma_{i,j} \Psi^{(i,j)}>0$.
We will consider the linear combination
\begin{align}
\Psi^{(1,3)}-c \gamma \Psi^{(0,1)}=\frac{\sqrt{\pi } c^4 \left((\delta-1) \delta \left((2 \delta-1)^3+135 \eta^2\right)+360 \pi  \gamma  \eta^2+15 \eta^2\right)}{155520 \Gamma \left(\frac{1}{4}\right)^2}+\mathcal{O}(c)^3,
\label{largecbnd2}
\end{align}
\begin{figure}
	\centering
	\includegraphics[scale=.6]{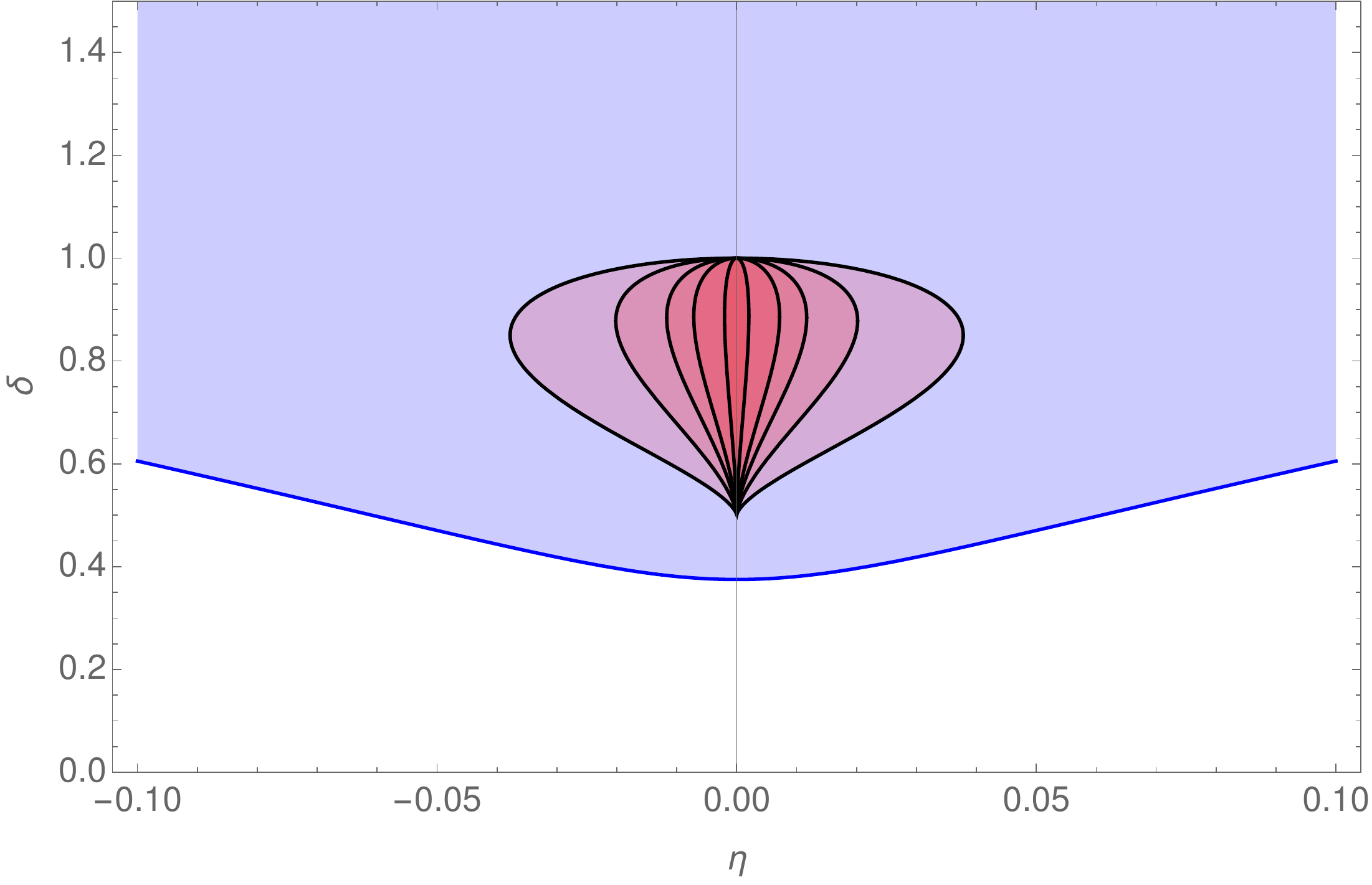}
	\caption{\label{fig2} The red regions are where $\Psi^{(1,3)}-c \gamma \Psi^{(0,1)}>0$; modular invariance implies the existence of an operator in each of these regions for any value of $\gamma$. The regions become narrower as $\gamma$ is increased. The blue region is the unitarity bound set by the positivity of the Kac matrix. This is a stronger constraint than the one in Figure \ref{fig1}.}
\end{figure}
where $\gamma$ is a constant. The regions where these are positive are plotted in Figure \ref{fig2}, as a function of $\gamma$.  The boundary of these regions is given by the curves
\begin{align}
\eta(\delta)=\pm\frac{i \sqrt{\delta-1} \sqrt{\delta} (2 \delta-1)^{3/2}}{\sqrt{360 \pi  \gamma +135 \delta^2-135 \delta+15}}.
\end{align}
In the limit where $\gamma$ becomes large the bounding region corresponds to a narrow slit in the $\eta$-$\delta$ plane bounded by $\frac{1}{2}<\delta<1$.  We conclude that every ${\cal W}_3$ CFT must possess an uncharged primary operator with $\frac{c}{24}<h<\frac{c}{12}$.

We have shown that modular invariance strongly constrains the spectrum of irrational ${\cal W}_3$ CFTs, but we have not obtained any results which are in contradiction with positivity of the Kac matrix.  It would be interesting to perform a higher order numerical analysis, and to consider separately the constraints obtained by acting with $\bar q$ derivatives, following \cite{Friedan:2013cba, Collier:2016cls}.  This would lead to more powerful constraints, which one might hope are in direct contradiction with unitarity. 

\section*{Acknowledgements}

We are grateful to C. Beem, S. Hellerman, Z. Komargodski, G. Ng, N. Paquette, I. Tsiares and X. Yin for useful conversations. A.M. and K.C. are especially grateful to Y. Gobeil for discussions on Kac matrices for other higher spin CFTs. T.H. is supported by the Simons Foundation, through the Simons Collaboration on the Non-perturbative Bootstrap. N.A.J. is supported by the National Science and Engineering Council of Canada and by the Simons Foundation, through the Simons Collaboration on the Non-perturbative Bootstrap.
K.C.~and~A.M.~are supported by the National Science and Engineering Council of Canada and by the Simons Foundation, through the Simons Collaboration {\it It from Qubit}.   T.H and A.M. acknowledge the Aspen Center for Physics, supported by the National Science Foundation grant PHY-160-7611,  where some of this work was performed. E.P. is supported by the Department of Energy under Grant No. DE-FG02-91ER40671.

\appendix
\section{No constraints from $\W_3$ higher-level Kac matrix for $c\leq98$}\label{app:w3level2}
In this appendix we prove that for $c\leq 98$, imposing unitarity of the $\W_3$ Kac matrix above level one yields no new bounds. 

Following \cite{Mizoguchi:1988vk, Mizoguchi:1988pf}, the determinant of the Kac matrix of $\W_3$ at level $L$ is of the form
\e{detmiz}{{\rm det} M_L = \prod_{k=1}^L \prod_{mn=k} (f_{mn}(h,c)-q_3^2)^{P_2(L-k)}\times({\rm positive})}
where
\e{}{\sum_{n=0}^\i P_2(n) q^n = \prod_{n=1}^\i{1\over (1-q^n)^2}}
and
\es{}{f_{mn}(h,c) = {64\over 9(5c+22)}&\left[h+(4-n^2)\a_+^2 + (4-m^2)\a_-^2 - 2+{mn\over 2}\right]\times\\&\left[h-4((n^2-1)\a_+^2+(m^2-1)\a_-^2)-2(1-mn)\right]^2}
with
\e{}{\a_{\pm}^2 = {50-c\pm \sqrt{(2-c)(98-c)}\over 192}}
Note that $f_{mn}(h,c)-q_3^2\in\mathbb{R}$ for $c\geq 98$. The first several $P_2(n)$ are
\e{}{P_2(n)  = \lbrace 1, 2, 5, 10, 20, 36, 65, 110, 185, 300, 481\rbrace}

First let's take $c=98$. The level-one bound is $h^2(h-3)\geq 72q_3^2$. For this value of $c$, $\a_+^2=\a_-^2$. This implies a symmetry,
\e{}{f_{mn}(h,98) = f_{nm}(h,98)}
Due to the structure of the $mn$ product in \eqr{detmiz}, the only terms that are not raised to even powers -- and can thus potentially give a negative determinant -- have $m=n$. This can only occur at levels $L=1,4,9,\ldots$. We have
\e{fmn}{f_{mm}(h,98) = \frac{1}{72} \left(m^2+h-4\right) \left(h+4 m^2-4\right)^2}
Noting that $f_{mm}$ increases with $m$, we see that $f_{mm}-q_3^2>0$ for any $m$ once the level-one constraint is satisfied. This implies that no new constraints will arise from $L>1$. 

Now let's take $c<98$. Now $\a_{\pm}^2$ are complex. But the structure of $f_{mn}(h,c)$ implies the following:
\e{}{c<98:\quad f_{mn}(h,c) = S_{mn}+i A_{mn}}
where 
\e{}{S_{mn}=S_{nm}~, \quad A_{mn}= -A_{nm}}
Thus,
\e{}{(f_{mn}-q_3^2)(f_{nm}-q_3^2) = (S_{mn}-q_3^2)^2 + A_{mn}^2 \geq 0}
Therefore, all of the $m\neq n$ terms give positive contributions. Finally, for any $c$,
\e{}{f_{mm}(h,c) = \frac{\left((c-2) m^2-c+24 h+2\right)^2 \left(96 h+(c-2)(m^2-4)\right)}{7776 (5 c+22)}}
Noting again that this increases with $m$, we see that terms $f_{mm}(h,c)-q_3^2\geq 0$ for all $m\geq 1$ after imposing the level-one constraint. This concludes the proof.

For $c>98$, the $m=n$ terms still give no constraint, but the off-diagonal terms are harder to analyze, due to the reality of $f_{mn}$. Experiments reveal nothing new; this is to be contrasted with the $\W_4$ case, as explained in footnote \ref{footnote:null}.

\sec{A sampling of irrational unitary CFTs with higher spin algebras, and their null states}
In this Appendix, we discuss some examples of unitary CFTs which are irrational with respect to a higher spin $\W$-algebra symmetry -- that is, $c>c_{currents}$. While we do not give a complete taxonomy,\footnote{A full classification of finitely-generated $\W$-algebras, much less a classification of irrational unitary CFT realizations thereof, is still absent. The literature on such CFTs is both vast and sporadic. Many early advances are summarized in the review of Bouwknegt and Schoutens \cite{Bouwknegt:1992wg}.} all such CFTs that we are aware of have the property that the vacuum module of $\W$ has non-trivial null states. The obvious examples are the higher spin minimal models, but it is true in other cases as well. This encourages the speculation in the Introduction. We will occasionally use the terminology ``generic'' to refer to algebras without non-trivial null states.

\sssec*{Vir$^2/\Z_2$} 
Consider taking a $\Z_2$ orbifold of two copies of any irrational Virasoro CFT, with central charge $c>1$.\footnote{We thank Simeon Hellerman for bringing our attention to this setup.} The resulting theory has central charge $2c$, $c_{currents}=2$, and a unique stress tensor. It also contains higher spin Virasoro primary currents. The simplest is a $s=4$ current, of the form %
\e{}{J_4(z) = \sum_{i,j=1}^2 c_{ij}:\!T_i(z) T_j(z)\!: + \sum_{i=1}^2 c_i\p^2 T_i(z)}
where the $T_i(z)$ are holomorphic stress tensors on the two copies, and the constants $c_i, c_{ij}$ are determined by the primary condition with respect to the new stress tensor $T(z) = T_1(z)+T_2(z)$. The theory also contains more currents, and many null states. 

We can see this by studying the vacuum character (e.g. \cite{Borisov:1997nc})
\e{}{\chi_{\vac}^{\Vir^2/\Z_2}(\t) = \half (\chi_{\vac}^2(\t) + \chi_{\vac}(2\t))}
where $\chi_{\vac}(\t)$ is the Virasoro vacuum character. For convenience, define
\e{}{F(\t) \equiv \prod_{n=1}^\i (1-q^n)^{-1} = q^{1/24}\eta^{-1}(\t)}
We now establish the following two facts. First, there are more higher spin currents than $T$ and $J_4$. And second, there are non-trivial null states. To see this, we may subtract the generic vacuum character for $W(2,4)$, the chiral algebra generated by $T$ and $J_4$ \cite{Blumenhagen1990, Kausch:1990bn}, from $\chi_{\rm vac}^{\Vir^2/\Z_2}$. The former is 
\e{}{\chi_{\vac}^{W(2,4)}(\t) = {(1-q)^2(1-q^2)(1-q^3)F^2(\t)}}
where the prefactors account for the trivial null states. Expanding at small $q$, one finds
\e{5.4}{\big(\chi_{\vac}^{\Vir^2/\Z_2}(\t) - \chi_{\vac}^{W(2,4)}(\t)\big)F^{-1}(\t)  = q^6+q^8+q^{10}+\ldots}
Using the generating function for the branching of some function $\chi(\t)$ into non-vacuum Virasoro primaries,
\e{}{\chi(\t)F^{-1}(\t) = \sum_{h}d_{p}(h) q^h~,}
where $d_{p}(h)$ is the number of Virasoro primaries at level $h$, \eqr{5.4} implies that we must add a $s=6$ Virasoro primary current. This cannot be a composite, so it is a generating element. We would still need to add (at least) this current if we had allowed null states in the algebra generated by $T$ and $J_4$, since that would have subtracted fewer states from $\Vir^2/\Z_2$. Now, because we have found three bosonic currents, and $3>c_{currents}=2$, there must be non-trivial null states. We can see them explicitly by subtracting the generic vacuum character of $W(2,4,6)$, the chiral algebra generated by currents of spins $s=2,4,6$, which gives
\e{}{\big(\chi_{\vac}^{\Vir^2/\Z_2}(\t) - \chi_{\vac}^{W(2,4,6)}(\t)\big)F^{-1}(\t)  = q^8-2q^{10}+\ldots}
The negative degeneracy indicates the presence of non-trivial null states. 

We note that there are many similar, but more intricate, examples of $\W$-algebras with and $c>c_{currents}$ that are formed via cosets. See e.g. \cite{Blumenhagen:1994wg}, or Section 7.3 of \cite{Bouwknegt:1992wg} for an earlier and less complete treatment.\footnote{Similarly, there are many such algebras formed via projection of generic algebras. One of the simpler ones is a non-generic $W(2,4,6)$ algebra constructed by projection of the $\N=1$ super-Virasoro algebra \cite{Bouwknegt:1988nz}.} These always contain non-trivial null states. 

\sssec*{Calabi-Yau sigma models} 
Consider sigma models on CY$_d$ manifolds. Such theories have $\N=2$ SUSY, and $c=3d$. In addition to the super-Virasoro generators, the current sector includes four Virasoro primary currents of spins $s=d/2, (d+1)/2$, two of each. Thus, the full algebra has $c_{currents}=6$. This algebra was constructed and studied by Odake \cite{Odake:1989ev,Odake:1989dm,Odake:1988bh}, who observed that for all $d$, there exist null states. The first ones occur at level one: if $J$ is the $s=1$ super-Virasoro current and $(G, G^\dag)$ are the $s=d/2$ currents, then 
\e{}{(J G)(z) = \p G(z)~, \quad (J G^\dag)(z) = -\p G^\dag (z)}
So for $d\geq 4$, these models contain higher spin currents, are irrational with respect to the above algebra, and contain non-trivial null states.

\sssec*{Sigma models on manifolds of $G_2,$ Spin(7) holonomy}
There exists a $c=12$ sigma model whose target space is a manifold of Spin(7) holonomy \cite{Shatashvili:1994zw}. The full chiral algebra is the so-called $SW(3/2,2)$ algebra, generated by the $\N=1$ super-Virasoro algebra together with a $s=2$ superconformal primary. Hence the set of Virasoro primary generating currents includes a higher spin current at $s=5/2$. The algebra has $c_{currents}=3$. Unitary realizations of $SW(3/2,2)$ lie in a discrete set, of which $c=12$ is a member, but contain non-trivial null states. These states, and the discreteness, may be understood as a direct consequence of the existence of a Virasoro subalgebra of $SW(3/2,2)$ which has central charge $c<1$. \cite{Gepner:2001px,Benjamin:2014kna}. Since irreducible highest weight representations that are unitary with respect to the full algebra must also be unitary with respect to all real subalgebras, the minimal model subalgebra (which is, at $c=12$, that of the Ising model) implies the above properties. 

The $G_2$ case is similar. There is a $c=21/2$ sigma model whose target space is a manifold of $G_2$ holonomy. The full chiral algebra may be obtained from the so-called $SW(3/2,3/2,2)$ algebra, which is generated by the $\N=1$ super-Virasoro algebra together with two superconformal primaries of $s=3/2,2$, modded out by an extra ideal \cite{Mallwitz:1994hh, Noyvert:2002mc, deBoer:2005pt}. This implies $c_{currents}=9/2$. Again, the set of Virasoro primary currents includes one with $s=5/2$. Unitary realizations of $SW(3/2,3/2,2)$ form a discrete set, of which $c=21/2$ is a member, but contain non-trivial null states on account of a minimal model subalgebra (which is, at $c=21/2$, that of the tricritical Ising model). 

\bibliographystyle{ssg}
\bibliography{WNBib}

\end{document}